\documentclass[conference]{IEEEtran}
\IEEEoverridecommandlockouts
\usepackage{url}
\usepackage{cite}
\usepackage{amsmath,amssymb,amsfonts}
\usepackage{algorithmic}
\usepackage{graphicx}
\usepackage{textcomp}
\usepackage{xspace}
\usepackage{xcolor}
\usepackage{array} 
\usepackage{multirow}
\usepackage{tikz}
\usepackage[hidelinks]{hyperref}

\def\BibTeX{{\rm B\kern-.05em{\sc i\kern-.025em b}\kern-.08em
    T\kern-.1667em\lower.7ex\hbox{E}\kern-.125emX}}

\begin{document}

\title{\huge Showcasing standards and approaches for cybersecurity, safety, and privacy issues in connected and autonomous vehicles}

\author{

\IEEEauthorblockN{Ricardo M. Czekster}
\IEEEauthorblockA{\textit{School of Computer Science and Digital Technologies, Aston University}\\
Birmingham, United Kingdom \\
r.meloczekster@aston.ac.uk}
}

\maketitle

\begin{abstract}
In the automotive industry there is a need to handle broad quality deficiencies, eg, performance, maintainability, cybersecurity, safety, and privacy, to mention a few.
The idea is to prevent these issues from reaching end-users, ie, road users and inadvertently, pedestrians, aiming to potentially reduce accidents, and allow safe operation in dynamic attack surfaces, for the benefit of a host of stakeholders.
This paper aims to bridge cybersecurity, safety, and privacy concerns in Connected and Autonomous Vehicles (CAV) with respect to Risk Assessment (RA) and Threat Modelling (TM) altogether.
Practitioners know the vast literature on this topic given the sheer number of recommendations, standards, best practices, and existing approaches, at times impairing projects and fostering valuable and actionable threat analysis.
In this paper we collate key outcomes by highlighting latest standards and approaches in RA and TM research to tackle complex attack surfaces as the ones posed by automotive settings.
We aim to provide the community with a list of approaches to align expectations with stakeholders when deciding where and when to focus threat related analysis in automotive solutions.
\end{abstract}


\section{Introduction}\label{s:introduction}
Adoption of Connected and Autonomous Vehicles (CAV) will invariably increase as road drivers select sustainable and smarter solutions given their obvious value-added features.
Unfortunately, these Cyber-Physical Systems (CPS) are susceptible to a host of cyber-attacks and vulnerabilities introduced due to fast-paced and strict time-to-market deadlines, which affects quality and reduce end-users' cybersecurity protections~\cite{sun2021survey}.

Generally, end-users of systems usually expect them to present almost no issues in terms of performance, security, or privacy.
These assurances are given by a plethora of ways, for instance, one might employ Threat Modelling (TM) which is the up-front evaluation of potential security issues affecting applications~\cite{xiong2019threat}.
They sit on top of larger Risk Management~\cite{iso31000} and governance activities to map security issues.
As a Risk Assessment (RA) technique~\cite{iec31010,nist2012guide,ross2022800}, TM sits together alongside Attack Trees~\cite{schneier1999attack,mantel2019meaning}, Bow Tie Analysis, or Monte Carlo Simulation (to mention a few)~\cite{iec31010}, as they are used to identify, analyse, evaluate, or communicate risks to broader audiences.
The ensemble of adopted techniques for hardening systems is crucial to point out design flaws to be fixed by software project teams when probing systems' entry points and potential weaknesses.
It is the job of the team to pick and choose the techniques that yield proper value to end-users given time, expertise, and budgetary constraints.

The contributions of this paper are:
\begin{itemize}
  \item An overview of the state-of-the-art of Risk Assessment techniques combined with Threat Modelling with focus on CAVs.
  \item A discussion and comparative analysis of RA and TM to address risk and its importance in CAV ecosystems.
\end{itemize}

This paper is organised as follows.
Section~\ref{s:rel-work} lists related work with a timeline of approaches directed at CAVs.
In Section~\ref{s:discussion} we discuss techniques and compare approaches and guidelines altogether.
We end our paper in Section~\ref{s:conclusion} with some insights into how to best align RA and TM in modern DevOps for the automotive industry and the inherent benefits that it can bring aimed at improved security posture and fewer defects reaching end-users in complex attack surfaces.

\section{Related Work}\label{s:rel-work}
Gritzalis et al. (2018)~\cite{gritzalis2018exiting} studied some RA methodologies outlining shared phases and addressing how each method computes risk.
Examples they have explored were: \textit{Expression des Besoins et Identification des Objectifs de S\'{e}curit\'{e}} (EBIOS), MEthod for Harmonized Analysis of RIsk (MEHARI), Operationally Critical Threat and Vulnerability Evaluation (OCTAVE) and variants (OCTAVE Allegro, OCTAVE-S), IT-Grundschutz, \textit{Metodolog\'{i}a de An\'{a}lisis y Gesti\'{o}n de Riesgos de los Sistemas de Informaci\'{o}n} (MAGERIT), Central Computing and Telecommunications Agency Risk Analysis and Management Method (CRAMM), Harmonized Threat Risk Assessment (HTRA), NIST.SP 800, RiskSafe, and CORAS, non-exhaustively.
Abouelnaga and Jakobs (2023)~\cite{abouelnaga2023security} discussed risk analysis methodologies for the automotive industry, comparing approaches.

There has been a steady interest in TM throughout the years with the publication of books by Swiderski and Snyder (2004)~\cite{swiderski2004threat}, Shoestack (2014)~\cite{shostack2014threat}, and by Tarandach and Coles (2020)~\cite{tarandach2020threat} focusing on real-world applications.
As TM methodologies we cite Spoofing, Tampering, Repudiation, Information Disclosure, Denial of Service, and Elevation of Privilege (STRIDE)~\cite{shostack2014threat}, the Process for Attack Simulation and Threat Analysis (PASTA)~\cite{ucedavelez2015risk}, LINDDUN~\cite{wuyts2020linddun}, Attack Trees~\cite{schneier1999attack,saini2008threat}, Persona non Grata, Security Cards, hTMM (Hybrid TM Method), Quantitative TMM, Trike, VAST (Visual, Agile, and Simple Threat) Modeling, INCLUDES NO DIRT, SPARTA, CORAS~\cite{lund2010model}, and other~\cite{xiong2019threat,tarandach2020threat}.

In the automotive industry the notion of performing Threat Analysis and Risk Assessment (TARA)~\cite{macher2016review,benyahya2023systematic} is central for understanding and communicating threats.
Luo et al. (2021)~\cite{luo2021threat} have surveyed the literature on TARA for connected vehicles with interesting discussions.
Additionally, Hazard Analysis and Risk Assessment (HARA) addresses functional safety and identifies potential hazards and issues in requirements.
Early attempts to automate threat models by Schaad and Borozdin (2012)~\cite{schaad2012tam2} focused on creating lightweight models suitable for application in early Software Development Life Cycle (SDLC).
Also looking at performing threat analysis in early SDLC stages, ie, in the architecture level, our previous research explored a mapping from architectural choices and their threat model correspondence~\cite{czekster2024inspecting} and continuous risk assessment capabilities in DevOps~\cite{czekster2024continuous}.

Threat analysis combined with RA has been discussed as early as 2013 by Ward et al. (2013)~\cite{ward2013threat}, showcasing the need for multiple standard alignment since the inception of cyber-security vehicular research.
Over the years, some attempts at TM surfaced such as ThreatGet~\cite{schmittner2020threatget,schmittner2021asset} that complies to ISO/SAE 21434~\cite{iso2021iso} in a model-based approach that employs automated risk identification whereas Hamad and Prevelakis (2020)~\cite{hamad2020savta} introduced SAVTA, a so called hybrid method for TM that generates attack trees.
Integrated threat modelling with risk analysis was investigated by Potteiger et al. (2016)~\cite{potteiger2016software} where authors provide a quantitative analysis leveraging the Common Vulnerability Scoring System (CVSS) metric.
Recently, it surfaced discussions and propositions on collaborative/cross-functional~\cite{von2022coretm} and automated~\cite{granata2024systematic} TM.

Amro et al. (2023)~\cite{amro2023assessing} employed MITRE's ATT\&CK framework~\cite{al2024mitre} for evaluating cyber-risk.
In terms of data sources to parametrise threat models, Jakstaite and Czekster (2023)~\cite{jakstaite2023extracting} collected threat intelligence data directly from social media outlets.
CAV are mobile and constrained CPS that involve road-drivers, passengers, pedestrians, and operators.
Techniques that analyse the validation and verification of such systems are translatable to CAV domains, eg, in formal modelling security~\cite{metere2024enhancing,yin2020recent,burmester2012modeling}, or cyber-attacks~\cite{lanotte2017formal,lian2024survey}.

\subsection{A brief timeline for RA and threat analysis in CAV settings}\label{ss:ra-vehicles}
Addressing risks in CAVs have a rich history combining a host of international institutes and researchers, as listed next.
\begin{itemize}
  \item \textbf{2009:} the E-Safety Vehicle Intrusion Protected Applications (EVITA)~\cite{henniger2009securing} project kickstarted a method for security risk assessment in automotive electrical/electronic (E/E) systems based on ISO/IEC 18045:2008 (this document was replaced by ISO/IEC 18045:2022~\cite{iso18045}).
  \item \textbf{2011:} publication of ISO 26262~\cite{iso26262:2011} discussing functional safety for road vehicles~\cite{debouk2019overview}.
  \begin{itemize}
    \item \textbf{2018:} Revision of the same document~\cite{iso26262:2018}, in $10$ parts, covering critical aspects.
  \end{itemize}
  \item \textbf{2014:} the National Highway Traffic Safety Administration (NHTSA), in the US, suggested the adoption of threat models for vehicular systems~\cite{mccarthy2014characterization} dubbed a \textit{composite modelling approach}.
  \item \textbf{2015:} The Security-Aware Hazard and Risk Analysis (SAHARA) method, by Macher et al. (2015)~\cite{macher2015}, aimed to combine concerns in security and safety altogether. It further develops ideas present in automotive HARA with STRIDE~\cite{potter2009microsoft,scandariato2015descriptive}.
  \item \textbf{2015:} definition of the Risk Analysis for Cooperative Engines (RACE)~\cite{boudguiga2015race} approach that combined EVITA and Threat, Vulnerability, and Risk Assessment (TVRA)~\cite{etsi-ts-102-165,etsi-ts-102-893}, the latter initially proposed by the European Telecommunications Standards Institute (ETSI) in 2011.
  \item \textbf{2016:} EVITA has heavily influenced other approaches such as HEAling Vulnerabilities to ENhance Software Security and Safety (HEAVENS)~\cite{olsson2016heavens} and Society of Automotive Engineers (SAE) J3061~\cite{sae-j3061,schmittner2016using}.  
  \item \textbf{2018:} proposition of the Security Automotive Risk Analysis (SARA)~\cite{monteuuis2018sara} method that considers safety, privacy, and security, offering a threat analysis framework, a mapping to attacks and assets, a modelling example using attack trees, and an observation metric.
  \item \textbf{2019:} suggestion of TARA+~\cite{bolovinou2019} model for cyber-security analysis of automated driving systems by performing functional safety. The approach was based on TARA and took into account remote attack surfaces (eg, modifications on infrastructure).
  \item \textbf{2019:} Maple et al. (2019)~\cite{maple2019connected} proposed a reference architecture suitable for attack surface analysis that included devices, edge, and cloud systems.
  \item \textbf{2021:} it was published the final draft of ISO/SAE DIS 21434~\cite{iso2021iso,macher2020iso} tackling cybersecurity for automotive domains that replaced SAE J3061.
  \item \textbf{2021:} publication of two United Nations (UN) regulations regarding cybersecurity (UNECE WP.29/R155)~\cite{UNECE155} and software updates (UNECE WP.29/R156)~\cite{UNECE156} that provided binding regulations for CAVs.
  \item \textbf{2022:} publication of ISO 21448:2022~\cite{iso21448:2022} on safety of road vehicles to ensure the Safety of the Intended Functionality (SOTIF).
\end{itemize}

Government, institutions, and the general community have joined forces and suggested the foundational frameworks on which vehicle manufacturers should follow as guidelines and recommended approaches to embed in their architectures for improved security posture.
However, from that point onwards, there is a need to check whether or not those bodies are in fact adopting these documents in their assembly lines, points we discuss in our next section.

\section{Discussion}\label{s:discussion}
In terms of importance and recognition throughout the years we highlight ISO 26262 (latest revision in 2018), ISO/SAE DIS 21434 (2021), and UNECE R155 and R156 (dated 2021) for handling security requirements in CAV, and ISO 21448 (2022), on safety.
These documents are the \textit{de facto} reference and guidance for addressing the most crucial security underpinnings when dealing with complex attack surfaces. 

It is noticeable that some documents focused on protocols for telecommunications and surrounding issues arising in CAV architectures, eg, the case of TVRA by ETSI.
Despite the importance of communications in any infrastructure and the inherent number of threats that exist only when one considers this dimension, there are other aspects that vehicular stakeholders must consider.
Next, we point out some issues to guide the focus of research and development of future vehicular networks for CAVs:

\begin{itemize}
  \item Combine security (cyber and physical), safety, and privacy altogether within a broader risk assessment phase~\cite{sion2019privacy}, in a holistic fashion, viewing and mapping assets for protection and identifying potential vulnerabilities whilst protecting user data.
  \item Understand the multiple layers on which risk, threats, and vulnerabilities sit, ie, in organisational levels, development, testing, and analysis.
  \item There are clear advantages on simplifying the processes involving risk and threat hunting across the board through modelling and abstractions, especially when communicating issues to non-technical stakeholders.
  \item Need to address changing attack surfaces in terms of infrastructure, mobility, and flexibility when approaching TM and other RA within any CPS/CAV contexts.
  \item STRIDE is a suitable method to be employed in early SDLC as it provides a framework to think about \textit{``what could go wrong?''} -- among other questions related to TM as described in the Threat Modeling Manifesto\footnote{Link: \url{https://www.threatmodelingmanifesto.org/}.} -- or how developers might go on attacking/abusing their own platforms and underlying systems/sub-systems. It provides stakeholders with good understanding and application of the CIA triad (Confidentiality-Integrity-Availability) amenable to diverse audiences.
  \item As in any security-safety-privacy effort, there should not exist a `one-size-fits-all' mindset, as systems must employ heterogeneity (when using operating systems and software libraries, etc) as means of defence. This recommendation does improve overall security where RA and TM are described in a generic enough fashion so it can be tailored to any environment posed by stakeholders.
  \item Whilst updating threat models to capture situations or events that have changed, or have been addressed in other SDLC phases, by the time one finishes updating it, it might be obsolete, which outlines the need for run-time generation of threat models in continuous RA~\cite{verreydt2024run}.
  \item Despite numerous efforts, organisations and researchers refrain from sharing their threat models, due to a host of reasons, the major one being exposure of security related issues to potential adversaries or competitors.
\end{itemize}

The sheer number of possibilities for addressing security in automotive settings, combined with strict time-to-market deadlines, adherence to regulations and industry recommendations, appropriate end-product quality specifications (where a reduced number of defects reaches end-users), among other issues, may cause confusion in teams working on solutions.
They must meet requirements that overlook security/privacy/safety concerns, under budgetary constraints, within limited time frames.
Table~\ref{t:comparative} aims to shed light on these issues by comparing most significant approaches.

\begin{table}
\centering
\caption{Non-exhaustive list of approaches/standards in CAV.}
\begin{tabular}{l|c|l}\hline
\multicolumn{1}{c|}{\textbf{Approach / Standard}} & \textbf{Year} & \multicolumn{1}{c}{\textbf{Focus}} \\ \hline\hline
EVITA & 2009 & Safety \\
STRIDE & 2009 & CIA/Software \\
TVRA & 2011 & Threat Analysis \\
$\rightarrow$ ISO 26262:2018 & 2011,2018 & Functional Safety \\
SAHARA & 2015 & Hazard and Risk \\
SAE J3061$\dagger$ & 2016 & Cybersecurity \\
HEAVENS & 2016 & Risk Assessment \\
SARA & 2018 & Safety, Privacy, Security \\
TARA+ & 2019 & Functional Safety \\
$\rightarrow$ ISO/SAE DIS 21434:2021 & 2021 & Cybersecurity \\
$\rightarrow$ UNECE R155 & 2021 & Cybersecurity \\
$\rightarrow$ UNECE R156 & 2021 & Software \\
$\rightarrow$ ISO 21448:2022 & 2022 & Safety \\ \hline
\multicolumn{3}{l}{\textbf{Legend}: `$\rightarrow$': standards.~~~ `$\dagger$: replaced by ISO/SAE DIS 21434:2021.}
\end{tabular}
\label{t:comparative}
\end{table}

Original Equipment Manufacturers (OEM) focus on adherence to standards (ISO 26262, ISO/SAE 21434, UNECE R155/R156, and ISO 21448) and then might use techniques and approaches (EVITA, SAHARA, TVRA, STRIDE, SARA, HEAVENS, TARA+, and so on) to help build more secure and safer solutions, embedded with data assurances.

We highlight also thinking systems in terms of \textit{data flows}, where data gets tracked and accounted for, at any stage, as these diagrams are invaluable for TM and adherence to privacy guarantees~\cite{myagmar2005threat}.
This process, known as `diagramming', is valuable to reason on potential security/privacy/safety violations in early specifications and during development is crucial for threat analysis.
It allows quick communication of issues to stakeholders, where they can devise prioritisation and severity analysis.
One could use for example Data Flow Diagrams (DFD)~\cite{shostack2008experiences,sion2018solution,sion2020security,tarandach2020threat} as a valuable (and simple) technique for tackling TM across SDLC steps.

\section{Conclusion}\label{s:conclusion}
In this paper we analysed approaches, standards, and methodologies converging at threat modelling and risk assessment and applicability in connected and autonomous vehicles research.
As outlined, we compiled key guidance and documentation that stakeholders in CAV should consult when developing, testing, and analysing the security posture of solutions.
We noticed that there is a shy integration for handling risks in CAV in established RA methodologies previously worked out by European institutions such as ISO or NIST/US.
The automotive community could profit from those outcomes to offer safer and more secure solutions to stakeholders.

As future work we shall investigate the intricacies of those different methods and attempt a more thorough comparison among approaches highlighting benefits and shortcomings.
We aim to outline specific gaps in threat analysis and how they cross-fertilise with modern SDLC approaches such as DevOps.



\bibliographystyle{ieeetr}
\bibliography{biblio}

\end{document}